\documentclass[twocolumn,showpacs,preprintnumbers,amsmath,amssymb, prb]{revtex4}
 
\usepackage{graphicx}
\usepackage{dcolumn}
\usepackage{bm}

\begin{document}


\title{Activation of additional energy dissipation processes  in the magnetization dynamics of epitaxial chromium dioxide films }

\author{G.~M. M\"uller}
 \email{mueller@ph4.physik.uni-goettingen.de}
\author{M. M\"unzenberg}%

\affiliation{%
IV. Physikalisches Institut, Universit\"at G\"ottingen, D-37077 G\"ottingen, Germany
}%

\author{G.-X. Miao}
\author{A. Gupta}
\affiliation{MINT Center, Department of Chemistry, Chemical and Biological Engineering, University of Alabama, Tuscaloosa, AL 35487
}%

\date{\today}

\begin{abstract}
The precessional magnetization dynamics of a chromium dioxide$(100)$ film is examined in an all-optical pump-probe setup. The frequency dependence on the external field is used to extract the uniaxial in-plane anisotropy constant. The damping shows a strong dependence on the frequency, but also on the laser pump fluency, which is revealed as an important experiment parameter in this work: above a certain threshold  further channels of energy dissipation open and the damping increases discontinuously. This behavior might stem from spin-wave instabilities. 
\end{abstract}

\pacs{75.30.Gw, 76.90.+d}
\maketitle

As a predicted half-metallic ferromagnet\cite{Schwarz1986}, Chromium dioxide ($\mathrm{CrO_2}$) shows a spin polarization at the Fermi level that comes close to a full polarization\cite{Dedkov2002, Anguelouch2001, Ji2001}, which is the defining property of a half-metal. Therefore, $\mathrm{CrO_2}$  has attracted a lot of  interest as a possible material for future spintronic devices.\cite{Pickett2001, Wolf2001} To achieve high processing speed in such devices, a fundamental insight into the magnetization dynamics of $\mathrm{CrO_2}$ is needed.

An all-optical pump-probe setup utilizing fs laser pulses\cite{Ju1998, Kampen2002}  (time-resolved magneto-optical Kerr effect, TRMOKE) allows investigation of the magnetization dynamics of ferromagnetic films in the time domain.  
In this setup, the laser induced demagnetization\cite{Beaurepaire1996, Koopmans2005} and the subsequent remagnetization\cite{Djordjevic2007} is accompanied by a change of the equilibrium direction of magnetization, which can be understood as a ps field pulse\cite{Jozsa2004} that leads to precessional motion according to the Landau-Lifschitz-Gilbert (LLG) equation\cite{Landau1965, Gilbert2004}
\begin{equation}
\frac{d}{dt}\mathbf{M}=-\gamma_0\ \mathbf{M}\times\mathbf{H}_{\mathrm{eff}}-\frac{\alpha}{M}\mathbf{M}\times\frac{d}{dt}\mathbf{M}, \label{eq:llg}
\end{equation}
where $\gamma_0=\mu_0|\gamma|$ and the dimensionless parameter $\alpha$ accounts for the damping of the magnetic motion. This damping term is derived by introducing isotropic Rayleigh-like energy dissipation. In general, the microscopic mechanism for damping does not obey these assumptions. Nevertheless, the precessional motion can still be described by an effective and possibly frequency dependent damping parameter $\alpha_{\mathrm{eff}}$. It has been shown that this parameter can be extracted from the precessional motion traced in an all-optical TRMOKE setup.\cite{Kampen2002,Djordjevic2006} In these experiments, the equilibrium magnetization is canted out of the film plane so that the laser induced demagnetization comes along with a change of the direction of the shape anisotropy field of the sample.
Zhang et al. have demonstrated \cite{Zhang2002} that the in-plane anisotropy of a $\mathrm{CrO_2(100)}$ film can be utilized in an optical pump-probe setup for the generation of an in-plane anisotropy field pulse that induces precessional motion. Recently, several TRMOKE experiments with a similar configuration were reported.\cite{Zhao2005,Talbayev2006,Rzhevsky2007,Liu2007}  Here, we present a systematic all-optical measurement of the precessional frequency and damping of  a 300\,nm $\mathrm{CrO_2(100)}$  film. The examined pump fluency dependence of the sample shows the opening  of an additional channel of energy dissipation at a sufficiently high perturbation from the equilibrium configuration. 

\begin{figure}[h!]
\includegraphics[width=\columnwidth]{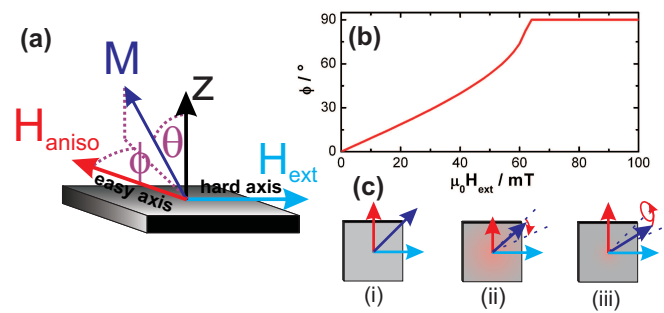}
\caption{\label{fig:config}(Color online) (a) Configuration of the sample system: The external field is applied in hard axis direction; the angles $\theta$ and $\phi$ are chosen according to Eq.~\eqref{eq:energy}. (b) $\phi$ calculated from Eq.~\eqref{eq:energy} with constants as given in the text. (c) In-plane anisotropy field pulse: the impinging pump pulse changes the equilibrium configuration (i) due to lattice heating and  adjustment of the magnetic anisotropy on the timescale of 1 ps (ii); the slow recovery of the intial configuration is accompanied by precessional motion (iii).}
\end{figure}

The $\mathrm{CrO_2}$ film is grown epitaxially by CVD on a $\mathrm{TiO_2(100)}$ (rutile) substrate.\cite{Miao2005}  
The 300 nm thick film, examined in detail for this work, is expected to show uniaxial magnetic in-plane anisotropy with the $c$-axis ($\left[001\right]$) being the in-plane easy axis and an effective first order anisotropy constant of $15\cdot 10^3\,\mathrm{J/m^3}$ at room temperature.\cite{Miao2005} This anisotropy originates from an interplay between crystalline and strain induced magnetic anisotropy. With the external field applied in the in-plane hard axis direction, the free energy density of the system can be written as
\begin{equation}\label{eq:energy}
F=-\mu_0M_{\mathrm{s}}H_{\mathrm{ext}}\sin\phi\sin\theta+\frac{1}{2}\mu_0M_{\mathrm{s}}^2\cos^2\theta+K_{1\mathrm{, eff}}\sin^2\phi,
\end{equation}
where the first term represents the Zeeman energy, contributions from the shape and the in-plane anisotropy are taken into account by the second and the third term, respectively. The angles $\phi$ and $\theta$ are named according to Fig.~\ref{fig:config} (a).

In our experiment, probe and pump pulse are generated by a Titanium:sapphire fs laser together with a regenerative amplifier (repetition rate 250 kHz, pulse width 60 fs, spot size of probe and pump pulse $40$ and $60\,\mathrm{\mu m}$, respectively). We utilize a double modulation scheme so that the polarization of the probe pulse is modulated with a photo-elastic modulator and the intensity of the pump pulse with a mechanical chopper.  The sample is mounted at room temperature with the external magnetic field in the plane of incidence parallel to the hard axis [cf. Fig.~\ref{fig:config} (a)]. It has been shown that the hysteresis curve for this sample in this configuration can be explained by coherent rotation of a single domain.\cite{Yang2000} Therefore, the equilibrium configuration, i.e., the azimuthal angle $\phi$, is given by the minimum of the free energy in Eq.~\eqref{eq:energy}. In Fig.~\ref{fig:config} (b), $\phi$ is depicted as a function of the external field where the room temperature material parameters are chosen to be $K_{1,\mathrm{eff}}=15\cdot 10^3\,\mathrm{J/m^3}$ and $\mu_0M_{\mathrm{s}}=603\,\mathrm{mT}$.\cite{Li1999}
\begin{figure}
\includegraphics[width=1\columnwidth]{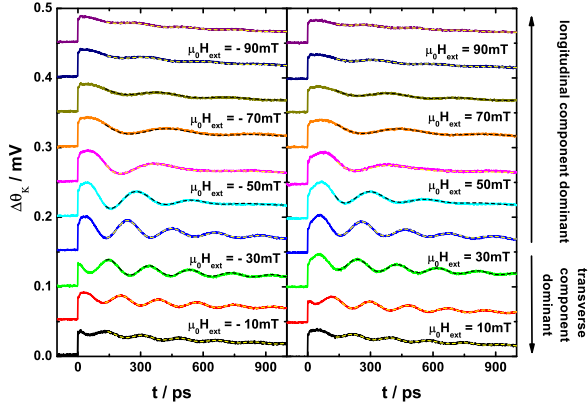}
\caption{\label{fig:transient}(Color online) Changes in the reflectivity and in the real part of the Kerr angle become manifest in the measured transients; the fits are given by the dashed lines. For lower fields, the transverse component of magnetization precesses dominantly where, for higher fields, the longitudinal component is the dominant one. The pump fluency is $F=10\,\mathrm{mJ/cm^2}$. The curves are shifted for clarity.}
\end{figure}

The laser induced demagnetization in $\mathrm{CrO_2}$ reaches its maximum after 200 to 300 ps -- in contrast to, e.g., nickel where the demagnetization occurs on a timescale of hundreds of fs. This behavior was attributed to the half-metallic character of chromium dioxide and a resulting decoupling between the spin and the electron system.\cite{Zhang2006, Muenzenberg2007} Nevertheless, the relevant time scale for electron-lattice equilibration is in the order of  1 ps in $\mathrm{CrO_2}$ as well as in nickel. Therefore, the magnetic in-plane anisotropy constant undergoes a fast change in magnitude on the same timescale, which  results in an anisotropy field pulse.  The original configuration recovers accompanied by precessional motion  on a timescale of 100 ps [cf. Fig.~\ref{fig:config} (c)]. 
\begin{figure}

\includegraphics[width=0.9\columnwidth]{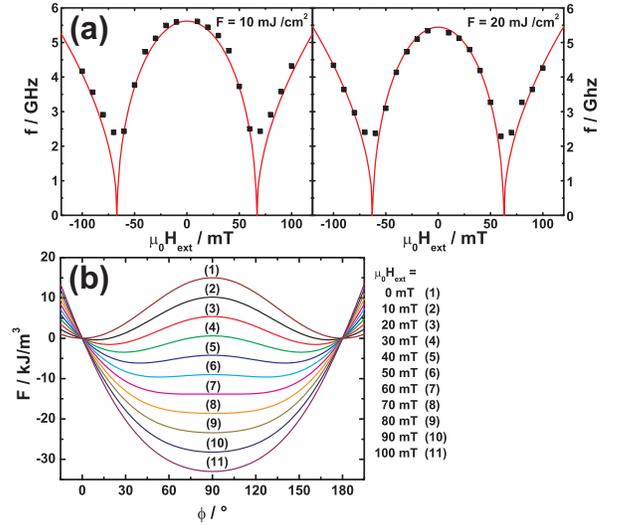}
\caption{\label{fig:freq}(Color online) (a) Measured precessional frequency as a function of the external field for pump fluencies $F=10\,\mathrm{mJ/cm^2}$ and $F=20\,\mathrm{mJ/cm^2}$ and fits according to Eq.~\eqref{eq:freq}. (b) Free energy density as a function of the azimuthal angle for different external field values.}
\end{figure}

The measured transients for pump fluency of $F=10\,\mathrm{mJ/cm^2}$ are depicted in Fig.~\ref{fig:transient}; it should be noted that these transients portray temporal changes in the real part of the Kerr angle as well as in the reflectivity. In the field range where the transverse component precesses dominantly ($40\,\mathrm{mT}<\mu_0 H_{\mathrm{ext}}<40\,\mathrm{mT}$, cf. Fig.~\ref{fig:transient}), the traced magnetic motion changes its phase by $\pi$ under switching the external field due to the the twofold degeneracy of the equilibrium configuration and the pecularities of the exciting field pulse. At higher fields where the longitudinal component is the dominant one, almost no change of phase is observed. Thus, a separation of the magnetic signal from the temporal change of the reflectivity would be complicated.
These transients can be perfectly fitted with functions of the type $A\cdot\exp(-t/T)+B\cdot\exp(-t/\tau)\cdot\sin(\omega t)$ where the first summand portrays the change in reflectivity and a small change in the length of the Kerr vector. The second one represents a solution of the linearized LLG equation [Eq.~\eqref{eq:llg}]. 
In the framework of this linearization, which corresponds to a parabolic approximation of the free energy, the precessional frequency $\omega=2\pi f$ and the damping time $\tau$ can be expressed by the sample parameters $M_{\mathrm{s}}$, $K_{1,\mathrm{eff}}$, $\alpha_{\mathrm{eff}}$, and the Land\'e factor $g$ as well as the external field $H_{\mathrm{ext}}$ (neglecting quadratic terms in $\alpha_{\mathrm{eff}}$):
\begin{eqnarray}
\omega&=&\frac{\gamma_0}{\mu_0M_s\sin\theta_0}\sqrt{F_{\theta\theta}F_{\phi\phi}-F_{\theta\phi}^2}, \nonumber\\
\tau&=&\frac{2\mu_0M_s}{\alpha\gamma_0\left(F_{\theta\theta}+\frac{1}{\sin^2\theta_0}F_{\phi\phi}\right)},
\label{eq:freq}
\end{eqnarray}
where $F_{ij}$ denotes the coefficients of the parabolic approximation of the free energy around the minimum.
In Fig.~\ref{fig:freq}(a), the extracted precessional frequency is plotted as a function of the external field  for pump fluencies of $F=10\,\mathrm{mJ/cm^2}$ and $20\,\mathrm{mJ/cm^2}$. The overall reduced frequency for the higher fluency can be explained by the higher average heating, which also manifests in a reduction of the Kerr signal at negative delay times. The measured frequencies are fitted by Eq.~\eqref{eq:freq}, where, to avoid ambiguities, the Land\'e factor and the magnetization are held fixed at $g=2$\cite{Lubitz2001}  and $\mu_0M_{\mathrm{s}}=603\,\mathrm{mT}$\footnote{Therewith, the total temperature dependence of the sample due to the average heating is included in the anisotropy constant which should exhibit the strongest temperature dependence.\cite{Callen1966}}. The determined values are $K_{1,\mathrm{eff}}=15,990(75)\cdot 10^3\,\mathrm{J/m^3}$ ($15,350(150)\cdot 10^3\,\mathrm{J/m^3}$) for $F=10\,\mathrm{mJ/cm^2}$ ($20\,\mathrm{mJ/cm^2}$).\footnote{The good agreement between measured and calculated frequencies also justifies the non-consideration of the fact that the length of the magnetization vector is not conserved in the experiment due to the slow laser-induced demagnetization. For all pump fluencies used in this work, the length of the magnetization vector (compared to negative pump probe delay times) is reduced by less than $4\%$.} The drop of the frequency at about $\mu_0 H_{\mathrm{ext}}=65\,\mathrm{mT}$ can be understood  easily: the precessional motion of magnetization means (in first order)  a  harmonic  oscillation around the minimum of the free energy landscape of the sample system [cf. Fig.~\ref{fig:freq}(b)].  At external fields below the anisotropy field $\mu_0H_{\mathrm{aniso}}=\frac{2K_{1,\mathrm{eff}}}{M_{\mathrm{s}}}$, there are two degenerate minima, which move towards each other with increasing field. At external fields above the anisotropy field, the magnetization is, of course, aligned with the external field. This transition from one to two minima at $H_{\mathrm{ext}}=H_{\mathrm{aniso}}$ must be accompanied by  vanishing second order derivatives at the minimum position and, thus, vanishing restoring forces for the precessional motion as well. 

\begin{figure}[h]
\includegraphics[width=0.9\columnwidth]{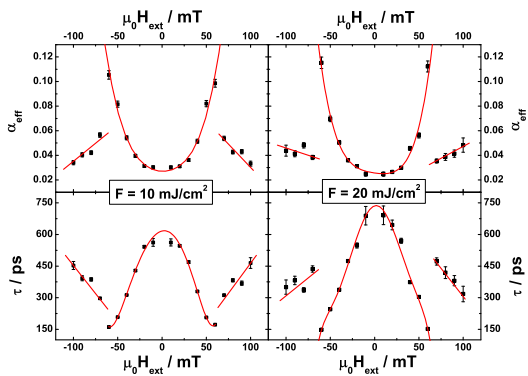}
\caption{\label{fig:damp}(Color online) Field depencene of the damping time and the effective Gilbert damping parameter for fluencies $F=10\,\mathrm{mJ/cm^2}$ and $F=20\,\mathrm{mJ/cm^2}$; the lines are guides for the eyes.}
\end{figure}

\begin{figure}[h]
\includegraphics[width=0.6\columnwidth]{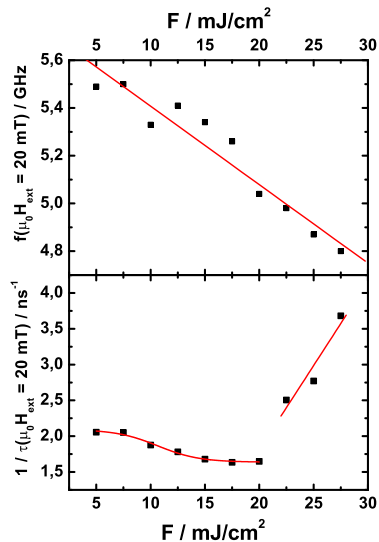}
\caption{\label{fig:fluency}(Color online) Pump fluency depencene of the precessional frequency and the inverse damping time for $\mu_0H_{\mathrm{ext}}=20\,\mathrm{mT}$; the lines are guides for the eyes. The energy dissipation rate increases above the threshold value of $F=20\,\mathrm{mJ/cm^2}$}
\end{figure}

Also, the damping time $\tau$ is extracted from the measured Kerr transients and plotted for the two pump fluencies in Fig.~\ref{fig:damp}. The damping parameter $\alpha_{\mathrm{eff}}$ is calculated from $\tau$ and the above determined sample parameters according to Eq.~\ref{eq:freq} and is included in Fig.~\ref{fig:damp}. The damping shows a strong increase with decreasing precessional frequency as it was also observed in similar experiments\cite{Zhao2005,Rzhevsky2007,Liu2007} so that genuine Gilbert damping cannot be the dominant process of energy dissipation operative in this sample. For CrO$_2$ films, an increase of the damping for lower precession frequencies was likewise found in ferromagnetic resonance (FMR).\cite{Lubitz2001} In addition, these FMR measurements revealed that the linewidth in the in-plane hard axis direction is broadened compared to  the easy axis direction. Thus, the damping in this sample might not only be frequency dependent, but also direction dependent. Woltersdorf and Heinrich could demonstrate that the magnetic damping  of an iron thin film on GaAs due to  two-magnon scattering is enhanced in directions of higher misfit dislocation density.\cite{Woltersdorf2004} In the $\mathrm{CrO_2}$ film, there might be a higher misfit dislocation density in the $b$-axis direction compared to the $c$-axis direction  due to the lattice mismatch anisotropy with the TiO$_2$ substrate.\cite{Gupta2006} Qualititatively, transmission electron microscopy reveals that the strain is relieved by formation of misfit dislocations that stretch to the surface of the film; a quantitative study of the supposed dislocation density anisotropy is not possible.  Another striking feature of the curves in Fig.~\ref{fig:damp} is the qualitatively different field dependence of the damping in the field range above $\mu_0H_{\mathrm{ext}}=60\,\mathrm{mT}$ for the two pump fluencies. Thus, the pump fluency is an important parameter in this all-optical measurement. In Fig.~\ref{fig:fluency}, the frequency and the inverse damping time $\tau^{-1}$ is plotted as a function of the laser fluency for an applied external field of $\mu_0H_{\mathrm{ext}}=20\,\mathrm{mT}$. The frequency decreases with the fluency as the equilibrium temperature rises due to the increased average heating.   In the room temperature regime up to 350 K, according to FMR results\cite{Lubitz2001}, the strength of damping is expected to decrease slightly with temperature. This behavior is observed in our experiment up to pump fluencies of $F=20\,\mathrm{mJ/cm^2}$; above this threshold, there is a discontinuous increase of the damping. By measuring the length of the Kerr vector in comparison with SQUID measurements, it is verified that the spin temperature remains below 330 K for all fluencies and time delays. Therefore, the increase of damping cannot be attributed to the increase of the spin temperature consistently with the FMR results. We suggest that this additional damping stems from spin wave instabilities that are long known to occur threshold-like in high-field-pumping FMR.\cite{Suhl1957} There, because of the high angle of excitation, the uniform precession becomes unstable due to spin wave disturbances and energy of the uniform mode is transferred to non-uniform ones. In time-resolved FMR where the precession is excited by a single field pulse, these instabilities have only been observed in very certain configurations.\citep{Silva2002, Gerrits2006} The question, which pecularities of the optically induced field pulse or the CrO$_2$ sample become manifest in the spin wave instabilities, suggested in this report, remains open for future research.

The authors gratefully acknowledge support by the Deutsche Forschungsgemeinschaft within the priority program SPP 1133.

\end{document}